\documentclass[12pt,preprint]{aastex}
\usepackage{natbib}
\def\etal{{\rm et al. }}
\def\a4{\hsize 17.0cm \vsize 25.cm}

\shorttitle{Hydrodynamic interplay between SMBH and nuclear starbursts}
\shortauthors{Hueyotl-Zahuantitla et al.}

\begin{document}

\title{On the Hydrodynamic Interplay Between a Young Nuclear Starburst and a Central Super Massive Black Hole}

\author{ 
Filiberto Hueyotl-Zahuantitla\altaffilmark{1},Guillermo Tenorio-Tagle\altaffilmark{1,2}, Richard W\"unsch\altaffilmark{3}, Sergiy Silich\altaffilmark{1} \& Jan Palou\v{s}\altaffilmark{3} }

\altaffiltext{1}{Instituto Nacional de Astrof\'\i sica Optica y
Electr\'onica, AP 51, 72000 Puebla, M\'exico; silich@inaoep.mx}
\altaffiltext{2}{Sackler Visiting Fellow, Institute of Astronomy, University of Cambridge, UK.} 
\altaffiltext{3}{Astronomical Institute, Academy of Sciences of the Czech Republic, Bo\v cni II 1401, 141 31 Prague, Czech Republic}

\begin{abstract}

We present 1D numerical simulations, which consider the effects of radiative 
cooling and gravity on the hydrodynamics of the matter reinserted by stellar 
winds and supernovae within young nuclear starbursts with a central 
supermassive black hole (SMBH). The simulations confirm our previous 
semi-analytic results for low energetic starbursts, evolving in a quasi-adiabatic regime, and 
extend them to more powerful starbursts evolving in the catastrophic cooling 
regime. The simulations show  a bimodal hydrodynamic solution in all cases.  
They  present a  quasi-stationary accretion flow onto the black hole, defined 
by the matter reinserted  by massive stars within the stagnation volume and a 
stationary starburst wind, driven by the high thermal pressure acquired in the 
region between the stagnation and the starburst radii. In the catastrophic cooling 
regime, the stagnation radius rapidly approaches the surface of the starburst 
region, as one considers more massive starbursts. 
This leads to larger accretion rates onto the SMBH and concurrently to powerful 
winds able to inhibit interstellar matter from approaching the nuclear starburst.
 Our self-consistent model thus  establishes a direct physical link between 
the SMBH accretion  rate and the nuclear star formation activity of the host galaxy  and provides a good upper limit to the accretion rate onto the central black hole.

\end{abstract}


\section{Introduction}

Powerful starbursts have been conclusively detected in the  nuclear regions of 
galaxies with active galactic nuclei (AGN; see for a review \citet{2005ARA&A..43..769V} 
and \citet{2009and..book..335H}). These include quasars \citep{2005ApJ...625...78H, 2008ChJAA...8...12H}, Seyferts 
\citep{2003ApJ...599..918I, 2007ApJ...671.1388D, 2008ApJ...677..895W, 2009ApJ...695L.130C}, 
submillimeter galaxies with an extreme star formation rate \citep{2005ApJ...632..736A, 2009Natur.457..699W} and even low luminosity AGNs \citep{1995MNRAS.272..423C, 2004ApJ...605..127G}. Some  supermassive black holes (SMBHs) coincide 
with massive and compact nuclear star clusters \citep{2008ApJ...678..116S}, some of them 
with a complex history of star formation are found in a number of nearby spiral 
\citep{2006AJ....132.1074R} and elliptical galaxies \citep{2006ApJ...644L..17W, 2006ApJS..165...57C}.

The interplay between young nuclear starbursts and their central SMBHs is not 
well understood and remains one of the central issue in  the theory of AGN galaxies, 
and in the process of cosmological growth of SMBHs \citep{2006MNRAS.368.1001L, 2009MNRAS.398...53B} and  their co-evolution with the bulges of their host galaxies 
\citep{2005MNRAS.361.1387B, 2005Natur.433..604D, 2008MNRAS.391..481S}. 
In fact, even basic issues regarding, for example, the impact of type II supernovae (SN) 
on the matter left over from star formation, seem to be  still undecided.   
One can find in the literature massive starbursts with a SN rate of 1 yr$^{-1}$ 
structuring a gaseous disk of just 10$^7$ M$_\odot$ \citep{2002ApJ...566L..21W} 
while other calculations assume that type II supernova might
evacuate most of the nuclear region from gas and dust \citep{2009MNRAS.393..759S} and
have considered only the evolution after the end of the type II SN epoch. 

\citet{2008ApJ...686..172S} presented a self-consistent, spherically symmetric stationary 
solution for the gaseous flow around a SMBH at the center of a young 
($\le 40$ Myr) starburst region. 
They also found a threshold mechanical luminosity ($L_{crit}$) which separates 
systems evolving in the quasi-adiabatic regime with a small stagnation radius, 
$R_{st}$  (at which the velocity is zero km s$^{-1}$) and consequently a low accretion 
rate,  from those with a large $R_{st}$ and a high accretion rate  whose hydrodynamics 
are strongly affected by  radiative cooling (see their Figure 6). 
The self-consistent semi-analytic solution found by \citet{2008ApJ...686..172S} is only valid  for 
nuclear starbursts with a mechanical luminosity $L_{NSB}<L_{crit}$. Above the threshold luminosity,  
strong radiative cooling drives the gas thermally unstable.   This has a strong impact on the 
thermal pressure gradient and thus on the  location of the stagnation radius, facts which 
inhibit a complete semi-analytic solution. Thus cases with a mechanical luminosity $L_{NSB}>L_{crit}$ 
demand of a numerical integration of the flow equations.
Nevertheless, from the semi-analytic results of \citet{2008ApJ...686..172S}  it was suggested  that 
the thermalization of the kinetic energy released within the volume occupied by a young  
nuclear starburst (NSB) should  lead to a central well regulated accretion flow 
onto the massive object, while  a powerful wind, capable of  preventing  the accretion of 
the ambient interstellar gas (ISM) onto the SMBH, should emanate from the outer regions. 
Here we follow these suggestions and measure the power of the resultant winds, 
as well as the accretion rates onto the SMBH and show how 
their luminosity  correlates with the starburst parameters.
Following \citet{2000MNRAS.311..346N},  \citet{2009ApJ...699...89C}
we have assumed that the angular momentum of the thermalized gas is not too large
and here we present results from our 1D numerical 
simulations as a first approximation. 
The simulations confirm Silich's \etal (2008) semi-analytic results  and 
extend them 
to starbursts with a mechanical energy input rate, $L_{NSB} > L_{crit}$.
In this paper  the input physics account for the  mechanical feedback that a young 
nuclear starburst may provide to the accretion flow while allowing  for  radiative cooling 
and the gravitational pull from both the central SMBH and the  nuclear starburst.  
The feedback provided by the central SMBH  \citep{1998A&A...331L...1S, 2001ApJ...551..131C, 2009ApJ...699...89C} will be the subject of a forthcoming communication. 

The paper is organized as follows: the numerical scheme used and the initial and 
boundary conditions are presented  in section 2. Section 3 deals with the results 
from the simulations and compares them  with the semi-analytic results. 
In section 4  the accretion rates and luminosities of the central SMBHs are obtained, 
as well as a measure of the possible impact of the resultant winds onto the host 
galaxy ISM. Section 5 gives our conclusions.

\section{Numerical Approach}

The numerical model is based on the finite difference Eulerian 
hydrodynamic code ZEUS3D v.3.4.2, which solves the set of the hydrodynamic 
equations \citep{1992ApJS...80..753S, 2008ApJ...683..683W}:
%------------------------------------------------------------------------ 
\begin{eqnarray}
      \label{eq1a}
      & & \hspace{-1.0cm}
\partial \rho/\partial t+\nabla\cdot(\rho u)=q_{m},
      \\[0.2cm]
      \label{eq1b}
      & & \hspace{-1.0cm}
\partial u/\partial t + (u\cdot\nabla)u + \nabla P/\rho=-\nabla\phi_{_{BH+NSB}},
      \\[0.2cm]
      \label{eq1c}
      & & \hspace{-1.0cm}
\partial e/\partial t + \nabla\cdot(eu)+P\nabla u=q_{e}-Q,
\end{eqnarray}
%-------------------------------------------------------------------------
where, $q_{m}$ and $q_{e}$ are the mass and energy deposition rates 
per unit volume, $e$ is the internal energy, $Q=n_{i}n_{e}\Lambda(T,Z)$ is
the cooling rate, $n_{i}$ and $n_{e}$ are the ion and electron number 
densities, and $\Lambda(T,Z)$ is the cooling function, which depends on the
thermalized gas temperature, $T$, and metallicity, $Z$. 
The Raymond \& Cox cooling function tabulated by \citet{1995MNRAS.275..143P} has been used
in all calculations. The right hand side of equation (\ref{eq1b}) represents 
the gravitational acceleration $a_{g} = -GM_r/r^{2}$, where 
$M_{r} = M_{BH} + M_{NSB}(r/R_{NSB})^{3}$ is the mass within a volume of
radius $r$. Outside of the starburst volume, $r > R_{NSB}$, it is 
$M_{r} = M_{BH}+M_{NSB}$.

All simulations have been carried out in spherical coordinates with symmetry 
along the $\theta$ and $\phi$ directions and with a uniform grid in the radial 
direction.  The calculations account for fast radiative cooling (see 
\citet{2007ApJ...658.1196T, 2008ApJ...683..683W}) and include the 
gravitational 
pull from the central SMBH and from the starburst,  assuming that the 
stars are  homogeneously distributed within a spherically-symmetric volume.  

\subsection{Initial and Boundary Conditions}

The initial distributions of velocity, pressure, temperature and density for 
all calculations were taken from the semi-analytic wind solution 
\citep{2004ApJ...610..226S}, without accounting for the gravitational pull from the  
starburst and the central SMBH. The initial condition is adapted to 
starbursts of the required size (see Table 1), 
with a mechanical luminosity $L_{NSB} < L_{crit}$. The mechanical luminosity 
$L_{NSB}=\dot M_{NSB}V_{A,\infty }^{2}/2$, (where, $V_{A,\infty }$ is the 
adiabatic outflow terminal speed, assumed to be 1500 km s$^{-1}$ in all 
calculations) has been normalized to the average mechanical luminosity for 
instantaneous starbursts  with a Salpeter initial mass function, sources 
between 1 M$_\odot$ and 100 M$_\odot$  and with ages less than 10 
Myr, $L_{NSB}=3\times10^{40}(M_{NSB}/10^{6}{\rm M}_{\odot})$ erg s$^{-1}$ 
\citep{1999ApJS..123....3L}.  The reinserted gas was assumed to have a negligible 
angular momentum and thus the flow could be solved in spherical symmetry. 

The mass and energy deposition rates per unit volume are 
$q_{m}=3\dot M_{NSB}/4\pi R^{3}_{NSB}$ and $q_{e}=3L_{NSB}/4\pi R^{3}_{NSB}$, 
respectively,  where $\dot M_{NSB}$ is the total mass deposition rate within 
the starburst volume (see Table 1). These values are added, at each time step,
to the computed density and total energy in every cell, $\rho_{old}$ and 
$e_{tot,old}$, respectively, when  located inside the starburst volume using 
the following procedure: $\rho_{new}=\rho_{old}~+~q_{m}dt$, and the velocity 
is corrected so that the momentum is conserved, 
$v_{mid}= v_{old}\rho_{old}/\rho_{new}$; 
the internal energy is corrected to conserve the total energy, 
$e_{i,mid}=e_{tot,old}~-~\rho_{new} v_{mid}^{2}/2$, and the new energy is 
inserted as a form of internal energy $e_{i,new}=e_{i,mid}~+~q_{e}dt$  (see, 
\citet{2007ApJ...658.1196T, 2008ApJ...683..683W}). 
The velocity of the flow at each radius is updated according to $ v_{new} = 
v_{mid}-a_{g}dt$, where $a_{g}=GM_{r}/r^{2}$ is the gravitational 
acceleration at each radius (see section 2).
The computational domain extends over the interval ($R_{in}, R_{out}$), where 
$0<R_{in}\ll R_{NSB} < R_{out}$. An open boundary condition was adopted at 
both ends of the computational grid. 

 %-------------------------------------------------------------------------
\begin{table}[!htbp]
\caption{\label{tab1} The input models}
{\small
\begin{flushleft}
\begin{tabular}{c c c c c c}
\hline\hline
Model&R$_{NSB}$ &$M_{NSB}$ &$\log~(L_{NSB})$& $L_{NSB}/L_{crit}$& $\dot M_{NSB}$\\
%\cline{7-9}
&{\scriptsize{(pc)}}&{\scriptsize{(10$^{8}$ M$_{\odot }$)} } & &  &{\scriptsize{(M$_{\odot }~{\rm yr^{-1}} $)}} \\
%\cline{6-8}
\scriptsize{(1)} & \scriptsize{(2)} & \scriptsize{(3)} &\scriptsize{(4)}
&\scriptsize{(5)} &\scriptsize{(6)}\\
\hline
1a&40&2.0&42.778&0.5      &8.45  \\
1b&40&4.0&43.079&1.0      &16.89 \\
1c&40&6.0&43.255&1.5    &25.34 \\
1d&40&6.8&43.302&1.7    &28.72  \\
1e&40&8.0&43.380&2.0    &33.79 \\  
2a&10&0.3&41.954&0.27    &1.27 \\
2b&10&0.5&42.176&0.45    &2.11 \\ 
2c&10&1.0&42.477&0.9    &4.22 \\
2d&10&1.67&42.698&1.5  &7.03  \\
2e&10&2.22&42.823&2.0     &9.38 \\
2f&10&2.75&42.916&2.5   &11.61 \\
2g&10&10.0&43.477&9.0    &42.23  \\
2h&10&20.0&43.778&18.0   &84.46  \\
\hline\hline
\end{tabular}
\end{flushleft}
     } %small
\footnotesize{The starburst input parameters for the simulations. Column 1 is a 
reference to the models. The radius ($R_{NSB}$), mass ($M_{NSB}$), logarithm of mechanical power ($L_{NSB}$, measured in erg s$^{-1}$),  the ratio of the starburst mechanical power 
to the critical mechanical luminosity ($L_{NSB}/L_{crit}$) and the total starburst 
mass deposition rate ($\dot M_{NSB}$) are presented in columns 2 to 6, 
respectively. All starburst models have a central SMBH with a mass  
$M_{BH} = 10^{8}$ M$_{\odot }$.} 
\end{table}
 %-------------------------------------------------------------------------

Our reference models are presented in Table 1. Here column 1 is a reference 
to the model, columns 2, 3 and 4 present the radius, mass and mechanical luminosity 
of the considered starburst, respectively. The ratio of the starburst mechanical 
luminosity to the critical luminosity and the total mass deposition rate inside the 
starburst volume are  shown in columns 5 and 6. The mass of the central SMBH, 
unless explicitly mentioned,  was assumed to be $M_{BH} = 10^{8}$ M$_{\odot}$ 
in all calculations. The computational domain for models 1a - 1e extends radially 
from $R_{in}=0.1$ pc to $ R_{out}=50$ pc. The inner and outer radii of the 
computational domain in the case of models 2a - 2h are 0.05 and 20 pc, respectively. 
1000 grid zones were used in all calculations.  The resolution convergency was tested 
in the case of the most energetic model (2h), carried out with 1000 and 3000 grid 
cells resolution. The results with both resolutions are in  excellent agreement over the whole 
computational domain.

\section{Results}
 
The results of the simulations are summarized in Table 2. Where, column 1 
is a reference to the model, column 2 presents the resultant stagnation radius. 
The total mass deposition rate is shown in column 3 and  should be compared  
with the resultant accretion rate and the mass outflow in the starburst wind, 
presented in columns 4 and 5, respectively. Column 6 shows the SMBH luminosity 
normalized to the Eddington limit ($L_{Edd}=1.3\times 10^{38}~M_{BH}$ 
M$_{\odot}^{-1}$ erg s$^{-1}$). The ram pressure of the outflow at the starburst 
edge ($P_{ram}=\rho u^{2}$) is presented in column 7.

%-------------------------------------------------------------------------
\begin{table}[!htbp]
\caption{\label{tab1} The predicted accretion rate and the power of the wind}
{\footnotesize
\begin{flushleft}
\begin{tabular}{c c c c c c c}
\hline\hline
Model& $R_{st}$& $\dot M_{NSB}$ & $\dot M_{acc}$ & $\dot M_{w}$&$L_{acc}/L_{Edd}$& $P_{ram}$\\
\cline{3-5}
& {\scriptsize{(pc)}} & \multicolumn{3}{c}{{\scriptsize{(M$_{\odot }~{\rm yr^{-1}} $)}}}& &{\scriptsize{(10$^{-7}$ dyn cm$^{-2}$)} }\\
%\cline{6-8}
\scriptsize{(1)} & \scriptsize{(2)} & \scriptsize{(3)} &\scriptsize{(4)}
&\scriptsize{(5)} &\scriptsize{(6)} &\scriptsize{(7)} \\
\hline

1a  & 2.1    &8.45   & $1.88\times10^{-3}$&8.44&$8.2\times10^{-4}$ & 1.96\\
1b  & 4.7    &16.89 & $4.33\times10^{-2}$& 16.86 &$1.9\times10^{-2}$ & 3.82\\
1c  & 11.4  &25.34 & 0.93 &  24.72 & 0.4 & 5.39\\
1d  & 14.4  &28.72 & 1.38 &  27.33 & 0.6 & 5.89\\
1e* & 17.5  &33.79 &2.88 &  30.89  & 1.2 & 6.57\\  
2a  & 1.3    &1.27   & $5.09\times10^{-3}$& 1.26 & $2.2\times10^{-3}$ & 4.66\\
2b  & 1.7    &2.11   &$1.02\times10^{-2}$& 2.07 & $4.5\times10^{-3}$  & 7.67      \\ 
2c  &2.8     &4.22   & 0.14 &  4.09   & $6.1\times10^{-2} $ & 14.64\\
2d  &4.4   &7.03   & 0.59 &  6.45   & $2.6\times10^{-1}$ &22.07\\
2e  & 5.3   &9.38   & 1.36 &  8.00   & 0.6  & 26.83\\
2f* & 5.9    &11.61 &  2.38 &  9.23  & 1.04  & 30.53\\
2g*& 8.2    &42.23 & 23.28 &18.74  & 10.2  & 56.16\\
2h*& 8.9   &84.46 & 59.54  &25.48  & 26.0  & 77.43\\

\hline\hline
\end{tabular}
\end{flushleft}
     } %small
\footnotesize{The predicted accretion rate and the power of the wind. Column 1 is a reference 
to the models. The results of the calculations: the value of the stagnation radius 
($R_{st}$), total starburst  mass deposition rate ($\dot M_{NSB}$), the calculated 
mass accretion rate ($\dot M_{acc}$), the rate at which matter flows away from 
the starburst as a super wind ($\dot M_{w}$), the stationary SMBH accretion 
luminosity normalized to the Eddington limit  ($L_{acc}/L_{Edd}$) and the ram 
pressure of the wind ($P_{ram}$) are shown in columns 2 to 7, respectively. 
Asterisks mark those models for which the calculated accretion luminosity exceeds the 
Eddington limit.}  %scritpstize
\end{table}
%-------------------------------------------------------------------------

\subsection{Comparison of the numerical and the semi-analytic solutions}

%-------------------------------------------------------------------------
\begin{figure}[!htbp]
\vspace{-2cm}
\includegraphics{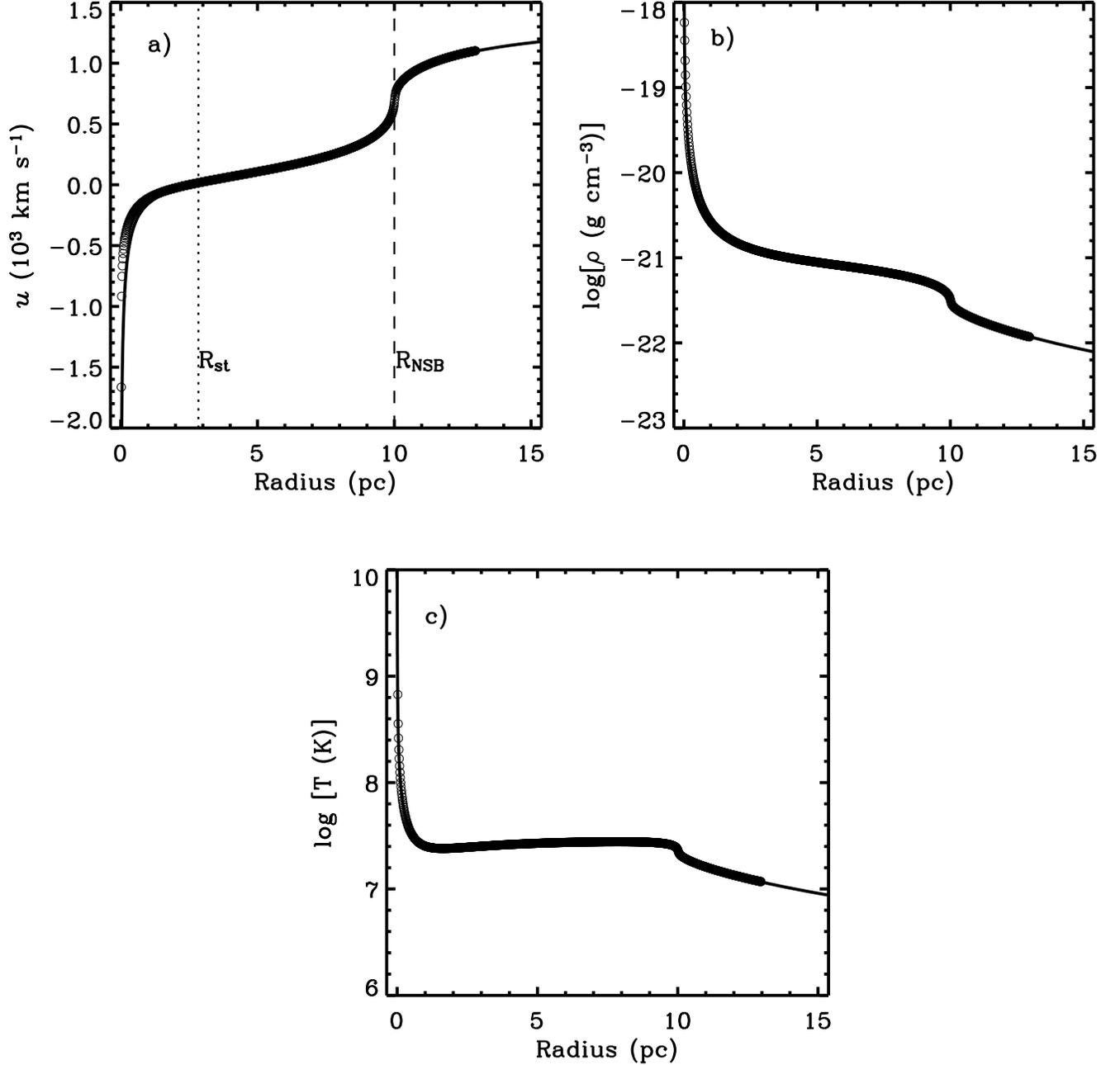}
\vspace{-3cm}
\caption{Test calculations. The numerical hydrodynamic solution (circles) for model 2c 
(see Table 2) is compared with the semi-analytic results (solid lines). 
Panels a-c  show the run of  the stationary velocity, density and temperature 
across the radial direction. In panel (a), the dotted and dashed vertical lines mark 
the location of the stagnation radius and the nuclear starburts radius, respectively.}
\end{figure}
%------------------------------------------------------------------------- 

In order to test our numerical code, several simulations were carried out for starbursts 
with $L_{NSB}  <  L_{crit}$ to compare them with the semi-analytic model of \citet{2008ApJ...686..172S}.  
Figure 1 presents, as an example, the results of the semi-analytic (solid line) and 
numerical (open circles) calculations for  case 2c in Table 1. There is  a good 
agreement between the two methods, as shown in panels a - c, for the stationary 
run of velocity, density and temperature, respectively. The value of the stagnation radius, 
marked by the  dotted line in panel (a), is $R_{st}=2.8$ pc, only about $\sim1.5$\% less 
than the value predicted by the semi-analytic model. The stationary solution shows 
how the matter deposited by massive stars inside the stagnation volume ends up 
falling towards the center and fuels the SMBH. On the other hand, matter reinserted 
between the stagnation radius and the starburst edge is steadily accelerated to 
reach its sound velocity at the starburst edge and then
it expands supersonically forming the starburst  wind. Note that as matter falls to the 
center its density grows orders of magnitude due to convergency alone (panel b). 
The temperature increases also very sharply (panel c)  due to the violent compression 
induced by the rapidly in-falling matter.

\subsection{The numerical solution above the threshold line}

%-------------------------------------------------------------------------
\begin{figure}[!ht]
\plotone{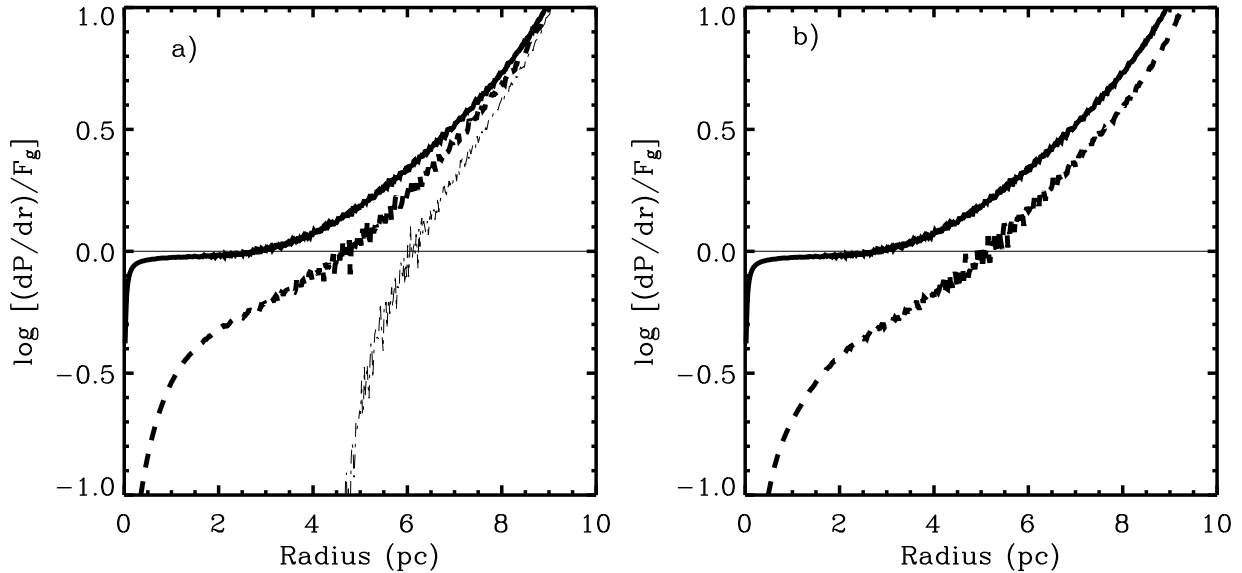}
\label{f2}
\vspace{-2cm}
\caption{The comparison of the pressure gradient to the gravity force inside 
the starburst volume. 
Panel (a) shows the ratio of the pressure gradient to the gravity force for starbursts of the 
same mass and radius (model 2c) but with different gas metallicities. Here solid, 
dashed and dotted lines correspond to $Z=Z_{\odot}$, $Z=5Z_{\odot}$ and 
$Z=10Z_{\odot}$, respectively.  
The intersection of the curves with the thin horizontal lines marks the position of the stagnation 
radius. The stagnation radius  moves to a larger distance from the center as the cooling rate 
becomes larger. Panel (b) shows the same ratio for starburst with different energy deposition 
rates (or different masses) but the same (solar) metallicity, the solid and dashed lines 
correspond to models 2c and 2e in Table 2, respectively.}
\end{figure}  
%-------------------------------------------------------------------------

The hydrodynamic solution for the matter reinserted by massive stars within an evolving  
young massive starburst in presence of a central SMBH is always bimodal, whether one 
considers starbursts  below or above the threshold line. The main difference is that above 
the threshold line strong radiative cooling becomes the physical agent that defines where the 
stagnation radius  lies.
The stationary location of the stagnation radius  is defined by the balance between the gravitational 
force ($F_{g}$) and the outward thermal pressure gradient  ($dP/dr$), which naturally, is 
strongly affected by energy  losses. Figure 2 presents the ratio of the pressure gradient to the 
gravity force as a function of distance from the center of the starburst. At the stagnation 
radius $dP/dr = F_{g}$ and thus the intersection of lines which display this ratio with the 
thin horizontal line marks the position of the stagnation radius for various cases. Figure 2a 
shows the ratio of the pressure gradient to the gravity force for starbursts with identical 
mass and radii (equal to those of model 2c in Table 1) when the thermalized gas was 
assumed to have  different metallicities. In these cases the displacement of the stagnation 
radius to larger and larger values is promoted by the increasingly larger amount of energy 
lost through radiative cooling within the considered starburst. 
Similarly, radiative cooling is enhanced  as one considers more massive starbursts. 
These reinsert more material per unit time and thus lead to a more significant  radiative 
cooling,  as shown in Figure 2b, for cases 2c and 2e.   In the region $r<R_{st}$, the outward 
pressure gradient is not able to compensate the gravity force and then all matter reinserted 
within the stagnation volume falls towards the central SMBH. On the other hand, in the 
region $R_{st}<r<R_{NSB}$ the pressure gradient exceeds the gravity force and hence the 
matter  deposited  there accelerates outwards and  conforms a supersonic wind.  
 Similar trends were noticed in the 3D results from \citet{2009MNRAS.393..759S} when considering 
the mass and energy input rate from planetary nebulae and type I SN in evolved clusters surrounding a SMBH.
Here however, we conclude that the amount of matter which fuels the central SMBH and that
which forms the starburst wind, both  depend directly on 
the location of the stagnation radius. 

%---------------------------------------------------------------
\begin{figure}[!htbp]
\vspace{-1cm}
\plotone{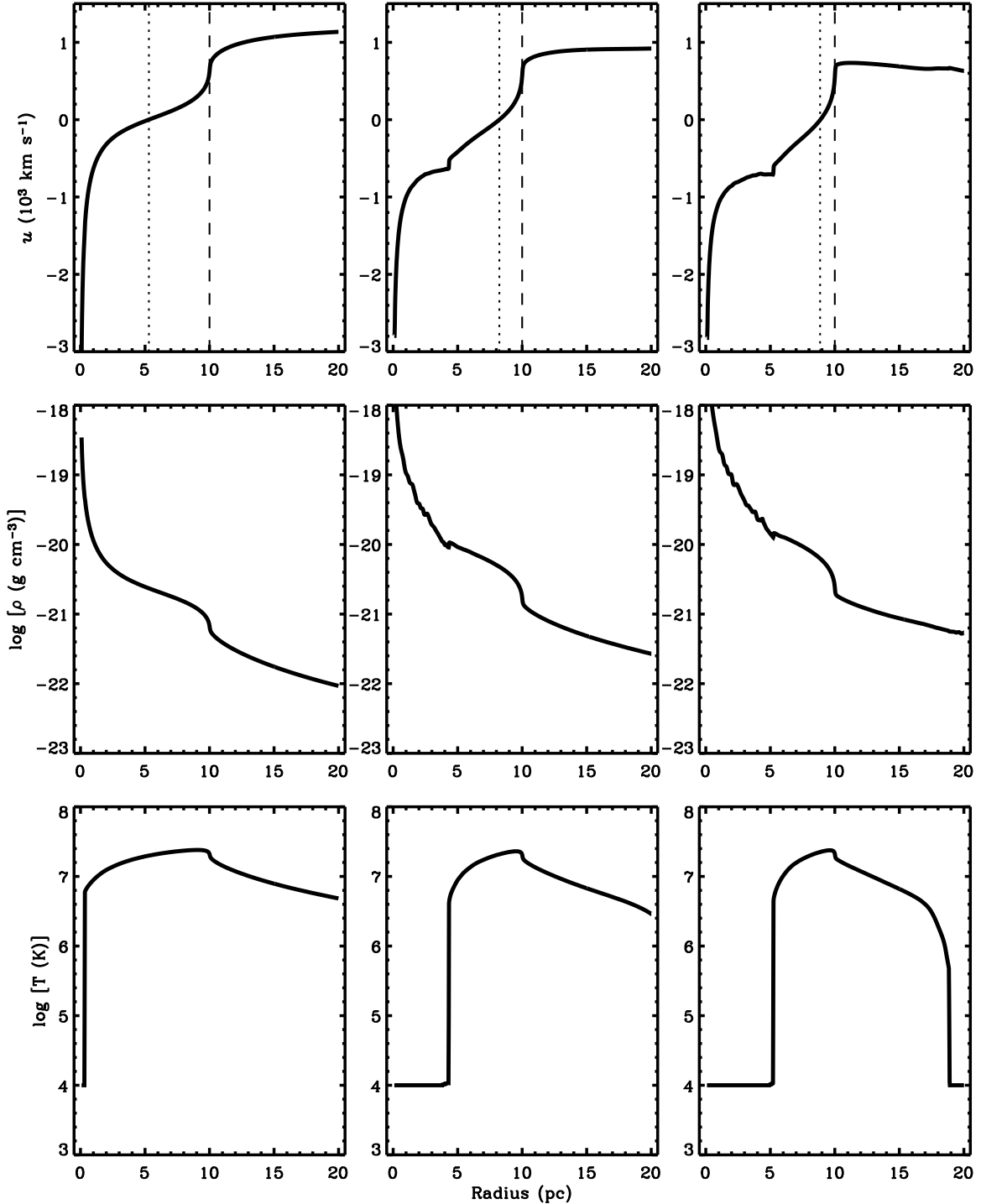}
\label{f3}
\vspace{-2.5cm}
\caption{{\small The hydrodynamic solution for starbursts with a mechanical 
luminosity 
$L_{NSB} > L_{crit}$ and a central SMBH. Panels from left to the right 
correspond to models 2e, 2g and 2h, respectively. Upper, middle and 
lower panels display the stationary velocity, density and temperature 
distributions, respectively. Dotted and dashed lines in the upper panels 
mark the location of the stagnation and the starburst radii, respectively.  
The stagnation radius appears further out for more energetic
(and thus more massive) starbursts, because the strong radiative 
cooling depletes rapidly the temperature to the minimum allowed in the 
calculations, $T_{min} = 10^{4}$ K,  and this results in a sudden lost of 
pressure. The radius of the thermally unstable zone also grows as the mechanical 
power of the considered starburst increases.}}
\end{figure}
%---------------------------------------------------------------

Figure 3 shows the results of  numerical simulations for starbursts with a 
mechanical luminosity larger than $L_{crit}$ (models 2e, 2g and 2h). Here 
the upper, middle and lower panels present the quasi-stationary distribution 
of the flow velocity, density and temperature, respectively.
As one considers more energetic (or more massive) starbursts, the larger  
densities (see middle panels) promote a faster radiative cooling within  the 
thermalized plasma and this results in a smaller pressure gradient and thus 
in a further displacement of the stagnation radius towards larger distances 
from the starburst center.  This is shown by vertical dotted lines in the upper panels.

The structure of the accretion flow for starbursts above the critical line presents 
some distinct features. In particular, the temperature distribution is different from 
that in the case of starburts below the threshold line. It drops smoothly within  the 
starburst  region until a thermal instability develops  within the accretion 
flow. The temperature then suddenly drops to the minimum permitted value 
as shown in the bottom panels in Figure 3.

Faster cooling leads also to a smaller wind speed. If one measures at a 
distance $r = 2R_{NSB}$ it is 1136 km s$^{-1}$ in the less energetic 
considered model (2e) and  629 km s$^{-1}$ in the most energetic case 
(model 2h), instead of  the 1500 km s$^{-1}$ expected in the adiabatic case.  
Note that for the most energetic case (model 2h), the temperature drops  
also suddenly to  $10^{4}$ K in the free-wind region.

Note that  despite the continuous input of energy, the gas temperature remains
then at the same value all the way, as it falls towards 
the center. Note also that  the thermal instability  appears  at larger distances from the 
center as one considers a larger mass deposition rate or a more massive and luminous 
starbursts (Figure 3, bottom panels). However, in all cases the cold and supersonically 
in-falling flow is well restricted to the central regions of the starburst, well within the 
stagnation volume. This is most probably the reason why the semi-analytic method is 
able to find with great accuracy the location of the stagnation radius, even 
for starbursts above the threshold line.   Figure 4 presents  the Jeans 
radius, $R_{J} = 0.5 c_{s}(\pi/G\rho)^{1/2}$ \citep{Clarke}, 
where $c_{s}$ is the local sound speed calculated at each $r$ inside the 
stagnation volume for our most massive models 2g and 2h. In both cases 
$R_J $ would be larger than $ r$ if the central SMBH is able to photoionize 
the accretion flow and the gas temperature cannot drop below $10^4$~K. 
However $R_J$ may be smaller than $r$, and thus the accretion flow may 
become gravitationally unstable if the temperature falls below $10^3$~K 
\citep{2009ApJ...702...63W}.  Note however, that for a given luminosity, the 
location of the stagnation radius is independent of the minimum temperature 
allowed in the flow and therefore, $\dot M_{acc}$ does not depend on the 
value of $T_{min}$, unless one quantifies star formation in the accretion flow, which is beyond the scope of the present study.  Thus, the accretion flow is always 
gravitationally stable in the quasi-adiabatic regime, below the threshold 
line, and becomes progressively more unstable above the critical line as one considers starbursts with a larger mechanical luminosity. 

%---------------------------------------------------------------
\begin{figure}[htbp]
%[!ht]
\plotone{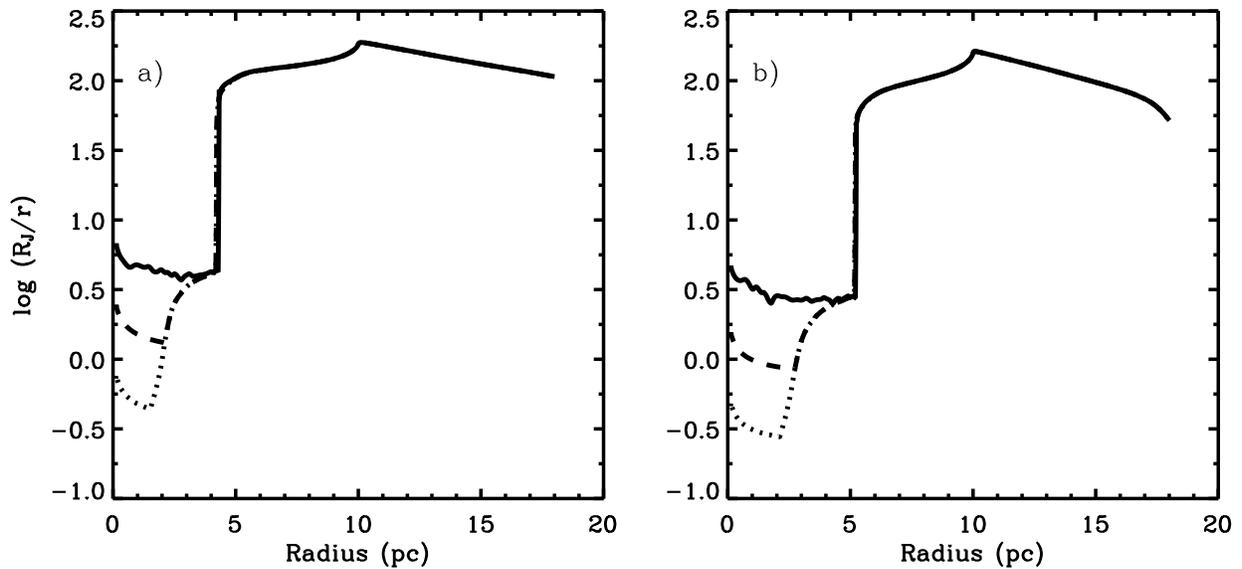}
\label{f4}
\vspace{-1cm}
\caption{The ratio of the Jeans radius to the radius of the flow as a function
of distance to the nuclear starburst center. Panels a and b present 
the results of the calculations for models 2g and 2h, respectively. The 
flow is gravitationally unstable if $R_J < r$. The solid, dashed and dotted
lines present log$(R_J/r)$ for different minimum temperatures allowed in the 
calculations: $T_{min} = 10^4$, 10$^3$ and 10$^2$~K, respectively.} 
\end{figure}
%---------------------------------------------------------------

The re-inserted matter thermalized inside the starburst region may 
contribute to the observed X-ray emission. The 0.2 - 8.0 keV X-ray 
luminosity  from the whole computational domain is (e.g. \citet{2005ApJ...635.1116S}): 
%---------------------------------------------------------------
\begin{equation}
      \label{eq1a}
L_x = 4 \pi \int_{R_{min}}^{R_{max}} r^2 n_e n_i \Lambda_X(T,Z) {\rm d}r 
\end{equation}
%--------------------------------------------------------------- 
where the electron and ion number densities are $n_{i} = n_{e} = 
\rho/\mu_{ion}$, $\mu_{ion}$ is the average mass per ion,
$\Lambda_X(Z,T)$ is the X-ray 
emissivity (see \citet{2000MNRAS.314..511S}), $R_{min}$ and $R_{max}$ are
the inner and outer boundaries of the computational domain. 
This is equal to about  2.5$ \times 10^{41}$ erg s$^{-1}$ and $2 \times 
10^{42}$ erg s$^{-1}$ for cases 2c and 2e, respectively.
The luminosity of the infalling matter is even smaller.
It is about $3.9 \times 10^{40}$ erg s$^{-1}$ and $1 \times 10^{42}$ 
erg s$^{-1}$ for cases 2c and 2e, respectively.
This emission is orders of magnitude smaller than the SMBH luminosity,
$L_{acc} = \eta_{acc} {\dot M_{acc}} c^2$, where $\eta_{acc}=0.1$ is 
the accretion efficiency and $c$ is the speed of light, which is: 
$L_{acc} \approx 8 \times 10^{44}$ erg s$^{-1}$ in case 2c and 
$L_{acc} \approx 8 \times 10^{45}$~erg s$^{-1}$ in case 2e.
And thus, it would be hard to 
quantify the infalling matter contribution to the total X-ray emission.
In this case, low luminosity AGNs \citep{2002ApJS..139....1T} seem to be  better 
candidates to show  the infalling matter X-ray emission, as in the 
case of the Seyfert 2/LINER galaxy NGC 4303. This shows the Raymond-Smith 
soft X-ray emission ($kT \approx 0.65$~keV) originating in the core of the 
galaxy with $r \le 15$~pc, coincident with a young (age around 4 Myr, and a 
3 pc radius), nuclear super star cluster (see \citet{2003ApJ...593..127J}).

\section{The starburst wind and the SMBH accretion rate and luminosity}

The hydrodynamic solution discussed in the previous section allows one to 
calculate the accretion rate and thus the SMBH luminosity for each model 
presented in Table 1. The time-dependent accretion rates for models 2c - 2f 
calculated as the mass flux through the inner grid boundary are shown as 
examples in Figure 5. At $t=0$ Myr the accretion rate, $\dot M_{acc}$, is 
equal to zero because a stationary wind solution with $R_{st} = 0$ was used 
as the initial condition (see section 2.1). However, the accretion rate grows 
rapidly and  reaches  an average value of 0.14 M$_{\odot }$ yr$^{-1}$, 0.59 
M$_{\odot }$ yr$^{-1}$,  1.36 M$_{\odot }$ yr$^{-1}$ and  2.38 M$_{\odot }$ 
yr$^{-1}$ for models 2c, 2d, 2e and 2f, respectively.  Note that, the 
accretion rate grows due to the larger stagnation volume and the larger mass 
deposition rate from more massive starbursts. After $\sim 0.1$ Myr the total 
mass is conserved $\dot M_{NSB} = \dot M_{acc}+\dot M_{w}$, see Table 2, 
columns 3, 4 and 5, respectively. Consequently, the fraction of the deposited 
mass expelled as superwinds from the starburst region decreases for more 
energetic starbursts although in absolute values it grows with the mass of 
the  considered  starburst (see Table 2).

%---------------------------------------------------------------
\begin{figure}[!ht]
\plotone{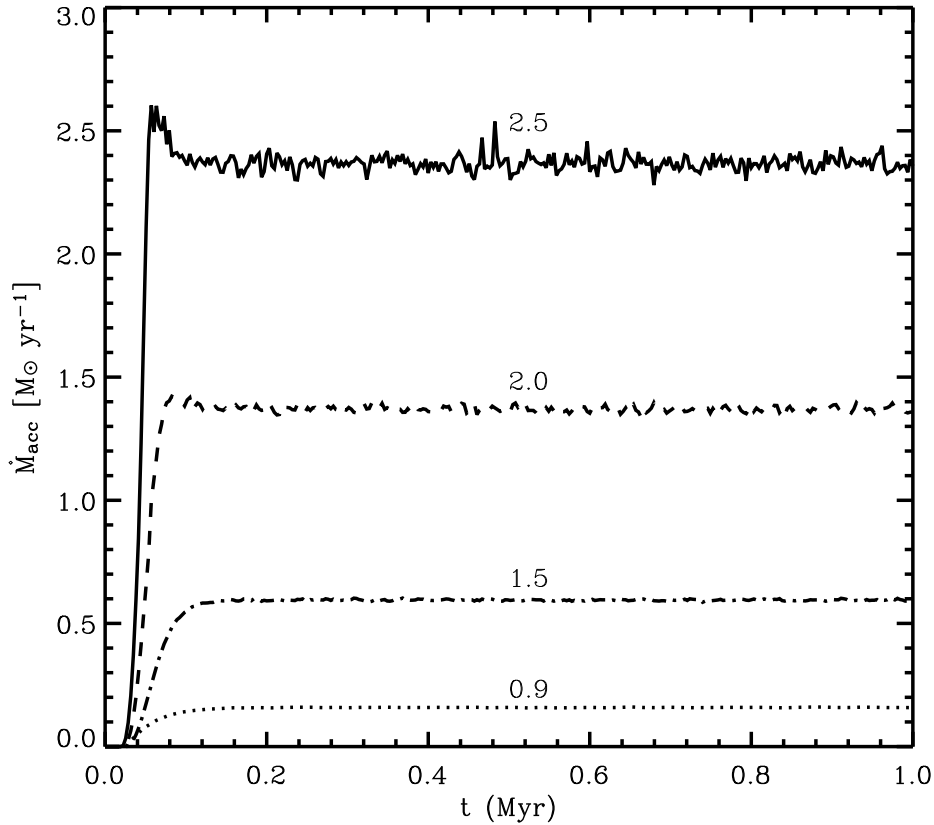}
\label{f5}
\vspace{-1cm}
\caption{ The time evolution of the mass accretion rate. The accretion rate onto the central $10^{8}$ M$_{\odot }$ black hole for models 2c, 2d, 2e and 2f; dotted, dash-dotted, dashed and solid  lines, respectively. The labels indicate the $L_{NSB}/L_{crit}$ values. Note that after $\sim 0.1$ Myr, $\dot M_{acc}$ remains almost constant and these values are reported in Table 2 together with $\dot M_{NSB}$ and the resultant $\dot M_{w}$. Note also that, the accretion rate associated to model 2f exceeds the Eddington limit. The small amplitude oscillations observed, mainly in models 2e and 2f, are numeric artifacts.}
\end{figure}
%---------------------------------------------------------------

The stationary accretion rates onto the SMBHs, $\dot M_{acc}$, and the stationary 
rate at which matter is ejected as a starburst wind, $\dot M_{w}$, obtained 
through the numerical integration of the flow equations and normalized to the 
total mass deposition rate, $\dot M_{NSB}$, are shown in Figure 6a as a function 
of the normalized starburst mechanical luminosity, $L_{NSB}/L_{crit}$. 
The circles represent the results from the numerical simulations  of models 
2a - 2e. The solid and dashed lines present the semi-analytic  $\dot M_{acc}$ 
and $\dot M_{w}$, respectively.  The mass accretion rate onto the SMBH grows 
more rapidly when the starburst mechanical power exceeds the critical value, 
($L_{NSB} > L_{crit}$). This leads to a rapid increase in the central SMBH 
luminosity, $L_{acc}=\eta_{acc} \dot M_{acc}c^{2}$, as shown in Figure 6b. 
There, the open circles result from our numerical simulations,  the solid and 
dotted lines present the semi-analytic results for starburst with $R_{NSB} = 
10$~pc (models 2a - 2e; see Table 1) and $R_{NSB}=$ 40  pc (models 1a - 1d), 
respectively. The cross symbols  mark the critical luminosity value ($L_{NSB} 
= L_{crit}$). Note, that the accretion rate and the SMBH luminosity obtained 
numerically are in a good agreement with those predicted by the semi-analytic 
model, even for starbursts with $L_{NSB} > L_{crit}$. This implies that the 
semi-analytic calculations lead to the correct value of the stagnation 
radius and thus may be used to estimate both the starburst wind power and the 
accretion onto the central SMBH and its corresponding luminosity, in all cases 
(above and below the threshold line). 

Note that in both sets of calculations ($R_{NSC}$ equal to 10 pc and 40 pc, with the assumed $V_{A,\infty}=1500$ km s$^{-1}$) the 
accretion rate reaches values $\sim 1.4$ M$_{\odot}$yr$^{-1}$ when $L_{NSB}\sim 2L_{crit}$. 
This could result in $\sim$50\% increase in the mass of the SMBH after $\sim$ 50 Myr. 
Note also that the calculated 
accretion luminosity exceeds the Eddington limit (see Table 2, models 1e, 2f, 2g, and 2h) 
when the starburst mechanical luminosity is just about twice its critical luminosity. 
Certainly, the accretion rate and hence the SMBH luminosity could be reduced if 
additional physics are included in the model. For example, one could think on 
a 2D or 3D geometry that could account for the radiative and/or mechanical 
feedback from the central AGN and the redistribution of the net angular  
momentum in the accretion flow (e. g. \citet{2009MNRAS.393..759S}). However, our 
1D model accounts for a realistic deposition of mass and energy 
around a central SMBH, and hence the results here presented
give a good estimate of the accretion rate upper limit. And more important of all, the model 
establishes a direct interplay between nuclear starbursts and their central SMBHs. A direct 
interplay in which all the re-inserted matter unable to join the superwind, becomes available to the SMBH.  The starburst wind, on the other hand, is sufficiently powerful in all cases, 
as to significantly re-structure the host galaxy ISM, leading perhaps to a thick 
ring, along the plane of the galaxy, and to a supergalactic wind along the host galaxy symmetry 
axis (as in \citet{1997ApJ...478..134T, 1998MNRAS.293..299T}). 

A simple estimate of the wind power can be obtained from its ram pressure ($P_{ram}$) at 
the starburst edge, see column 7 in Table 2. This is,  in all cases,  many orders of magnitude  
larger than the typical ISM pressure in our Galaxy ($\sim$ 10 $^{-12}$ dyn cm$^{-2}$). 
It also exceeds by almost three orders of magnitude  the pressure exerted by a one particle 
per cubic centimeter ISM, freely falling onto the starburst   ($P_{ISM}=\rho_{_{ISM}}v_{ff}^{2}$) 
with  $v_{ff}=[2G(M_{BH}+M_{NSB})/R_{NSB}]^{1/2}$.  The implication is thus that the resultant  
winds are to lead to the build up of superbubbles and probably to supergalactic winds,
preventing, in most cases, the falling of the ISM onto the nuclear starburst. Perhaps only in 
the case of  an extremely dense ISM ($\rho_{ISM} \sim 10^{-20} - 10^{-21}$ g cm$^{-3}$) 
freely falling onto the central starburst would modify the structure of the outflow.

%---------------------------------------------------------------  
\begin{figure}[!htbp]
\plotone{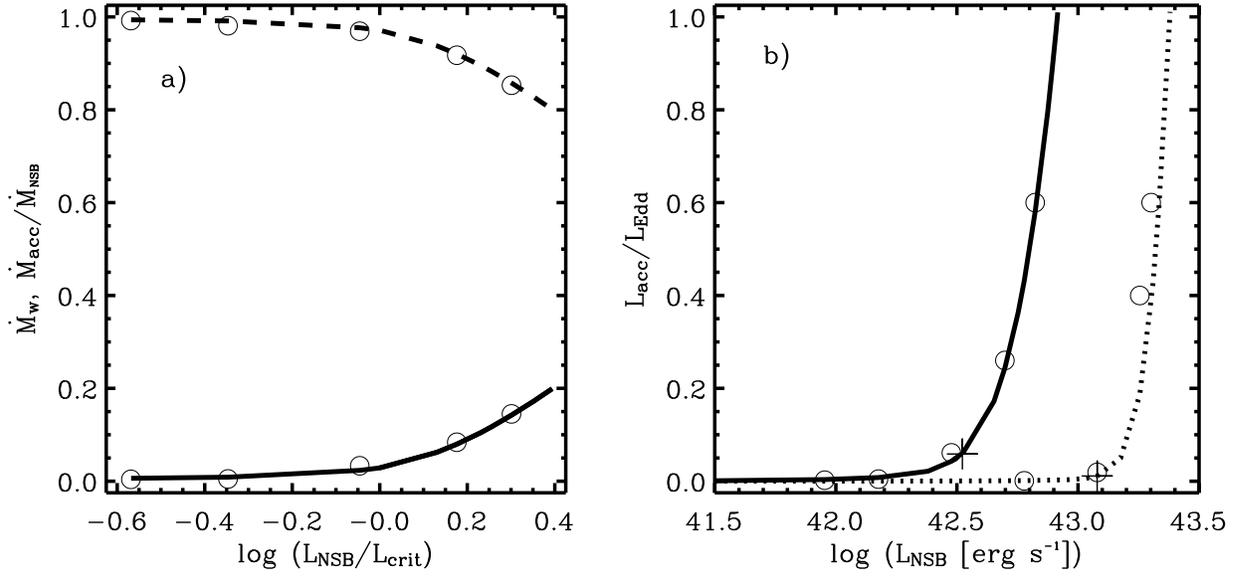}
\label{f6}
\vspace{-2cm}
\caption{The semi-analytic and numerical predictions for the accretion rate and the SMBH luminosity. 
Panel (a) presents the calculated mass accretion rate and the rate at which mass is expelled  as a 
superwind (circles) for starbursts below and above the threshold line (models 2a - 2e).  These are  
compared with the semi-analytic predictions (solid and dashed lines, respectively).  All rates have 
been normalized to the total starburst mass deposition rate, $\dot M_{NSB}$.   Panel (b) shows the 
SMBH luminosity normalized to the Eddington limit. The circles represent results from the numerical 
simulations, solid and dotted lines show semi-analytic calculations for starburst with 10 pc 
(models 2a - 2e) and 40 pc (models 1a - 1d), respectively. An accretion efficiency of $\eta_{acc}=0.1$ 
was used in the calculations.  Cross symbols  represent starbursts with  the critical energy ($L_{crit}$) input rate.} 
\end{figure}

\section{Conclusions}

By means of 1D numerical simulations and  semi-analytic estimates, we  have worked 
out the stationary hydrodynamic solution for  the matter  reinserted by stellar winds and type II supernovae from
a young, massive 
and compact starburst in presence of a  central SMBH.  The solution is bimodal in all cases, below and above the threshold line ($L_{crit}$), with a stagnation radius ($R_{st}$) which 
defines the outer boundary of the  accretion flow onto the SMBH  
as well as the inner boundary of the  starburst wind.

We have shown that at the stagnation radius, the force of gravity perfectly balances the 
outward pressure gradient  acquired by the thermalized reinserted matter.   We have 
also shown  that radiative cooling becomes an important issue for massive  starbursts 
with a mechanical luminosity above the threshold line ($L_{NSB} > L_{crit}$). In all these 
cases,  radiative cooling depletes the pressure established through  thermalization of the 
injected matter and this leads to the development of  a thermal instability in the accretion flow.  
The stagnation radius then moves rapidly, towards the starburst boundary, with the mass of the 
considered starburst. In all simulations with $L_{NSB} > L_{crit}$, strong 
radiative cooling occurs at a radius interior to the stagnation  radius. 
Radiative cooling re-structures the inner accretion flow lowering the 
temperature to the lowest allowed value, $T = T_{min}$. From then onwards 
and despite the continuous input of energy, the rapid velocity increase 
leading to a rapid density enhancement keeps and sustains the in-falling gas 
temperature at  $T = T_{min}$. It is the larger mass deposition rate,  
provided by more massive starbursts, what triggers the onset of strong  
radiative cooling and with it the shift of the stagnation radius  towards 
the starburst boundary.  This results in a rapid increase  of the central 
SMBH luminosity for starbursts further above the  critical threshold  in the 
$L_{NSB}-R_{NSB}-M_{BH}$ parameter space.

The larger mass deposition rates provided by more massive starbursts also leads 
to more powerful starburst winds  and an estimate of their mechanical power rules 
out  the possibility  of the ISM feeding the SMBH, at least during the type II supernova era.

Clearly, spherically symmetric calculations, as the ones presented  here, cannot account 
for the redistribution of the net angular  momentum in the accretion flow. Nevertheless, 
they provide a good  estimate of the upper limit to the accretion rate onto the central  black hole, 
while pointing to a direct  physical link between nuclear  starbursts and the central SMBH luminosity.  
Our calculations do  realistically account for the symmetric deposition of mass and energy from 
massive  stars around the central object. This suggests  that in a more  realistic 2D or 3D geometry, 
able to account for the redistribution of the net angular momentum, the hydrodynamics would still 
lead to a bimodal solution with an accretion flow and an outward wind. In such a case however, 
the residual angular momentum could favor the formation of a gaseous disk well contained within 
the nuclear starburst region. We shall consider some of these issues in a future communication.

\acknowledgments
This study has been supported by CONACYT - M\'exico, research grants
60333 and 82912 and the Spanish Ministry of Science and Innovation  under the
the collaboration ESTALLIDOS (grant
AYA2007-67965-C03-01) and Consolider-Ingenio 2010 Program grant
CSD2006-00070: First Science with the GTC. We also acknowledge the  CONACYT (M\'exico) and the  Czech Academy of Science
research grant 2009-2010, the institutional Research Plan AVOZ10030501 of the Academy of Sciences of the Czech Republic, and the project LC06014 - Center for Theoretical Astrophysics of the Ministry of Education, Youth and Sports of the Czech Republic. \citet{2005ApJ...632..736A}

\bibliography{fhz}
%\end{thebibliography}
\end{document}